\documentclass[%
 aip,
 apl, 
 reprint,
]{revtex4-1}

\usepackage{amsmath} 
\usepackage{amssymb}

\usepackage{graphicx}
\usepackage{dcolumn}
\usepackage{bm}

\usepackage[utf8]{inputenc}
\usepackage[T1]{fontenc}
\usepackage{mathptmx}
\usepackage{etoolbox}
\usepackage{multirow} 
\usepackage{float}

\makeatletter
\def\@email#1#2{%
 \endgroup
 \patchcmd{\titleblock@produce}
  {\frontmatter@RRAPformat}
  {\frontmatter@RRAPformat{\produce@RRAP{*#1\href{mailto:#2}{#2}}}\frontmatter@RRAPformat}
  {}{}
}%
\makeatother
\usepackage[colorlinks=true, linkcolor=blue, citecolor=red, urlcolor=magenta]{hyperref}

\newcommand{\Cltwoplus}{$\mathrm{(Cl_{2})}^{2+}$}
\begin{document}

\preprint{AIP/123-QED}

\title{A Third-Order Relativistic Algebraic Diagrammatic Construction Method for Double Ionization Potentials: Theory, Implementation, and Benchmark}
\author{Sujan Mandal}
\affiliation{Department of Chemistry, Indian Institute of Technology Bombay, Powai, Mumbai 400076, India}

\author{Achintya Kumar Dutta*}
\affiliation{Department of Chemistry, Indian Institute of Technology Bombay, Powai, Mumbai 400076, India}
\email{achintya@chem.iitb.ac.in}

\begin{abstract}
We present a relativistic third-order algebraic diagrammatic construction (ADC(3)) approach for calculating double ionization potentials (DIPs). Inclusion of third-order terms significantly improves the performance of the algebraic diagrammatic construction method for DIPs. By employing the exact two-component atomic mean-field (X2CAMF) Hamiltonian in combination with a Cholesky decomposition (CD) representation of two-electron integrals and the frozen natural spinor (FNS) framework for virtual space truncation, we achieve a significant reduction in both memory requirements and computational cost. The DIPs obtained using the X2CAMF Hamiltonian show excellent agreement with results from fully relativistic four-component calculations. We have validated the accuracy of our implementation through comparisons with available experimental and theoretical data for inert gas atoms and diatomic species. The effect of higher-order relativistic corrections is also explored.
\end{abstract}

\maketitle

\section{\label{introduction}Introduction}
Recent developments in experimental techniques, such as multi-electron detection using time-of-flight (TOF)\cite{eland2003complete} spectrometry and related photoelectron-photoelectron coincidence (TOF-PEPECO) spectroscopy,\cite{eland2003completea} have enabled detailed investigation of multiple ionization events caused by the absorption of a single photon. Moreover, cutting-edge free-electron laser (FEL) facilities, like the Linac Coherent Light Source (LCLS), now allow for the exploration of both sequential and direct multiphoton–multielectron ionization dynamics.\cite{young2010femtosecond,fang2010double} A pioneering study by Cederbaum et al.\cite{cederbaum1986double} predicted that creating double core holes at different atomic sites in a molecule offers an enhanced probe of local chemical environments—an idea that has now been experimentally validated.\cite{berrah2011doublecorehole} Similarly, Auger electron spectroscopy provides unique insights into a molecule’s electronic structure that are not accessible through photoelectron spectroscopy.\cite{carlson2013photoelectron,rye1984molecular} Since the Auger process leads to a final state with two fewer electrons, it is crucial to accurately model the resulting doubly ionized system to interpret these spectra reliably. Over the years, significant efforts have been devoted to calculating double ionization potentials (DIPs) using a variety of methods. In particular, methods based on the two-particle propagator or Green’s function formalism have been extensively utilized to evaluate the DIPs of atomic and molecular systems.\cite{liegener1982auger,liegener1983calculations,liegener1996auger,ortiz1984qualitative,tarantelli1985greens,tarantelli1987theoretical,ohrendorf1989doubly,graham1991multiconfigurational,tarantelli1993foreign,tarantelli1994aggregation,tarantelli1996greens,griffiths1998vertical,noguchi2005firstprinciples,ida2008secondorder,velkov2011intermediate} The Algebraic Diagrammatic Construction (ADC) theory, an efficient approximation to the two-particle propagator method, has been widely adopted in these studies, most often in its second-order form (ADC(2)). To the best of our knowledge, only one prior study has reported the implementation of the DIP-ADC(3) method in a non-relativistic framework is by Thielen et al.\cite{thielen2023development} In addition to the ADC framework, several studies have employed the equation-of-motion coupled-cluster (EOM-CC) variants for computing DIPs in the non-relativistic regime.\cite{nooijen1997similarity,nooijen2002state,sattelmeyer2003use,demel2008application,musial2011multireference,shen2013doubly} A recent study by Gururangan et al.\cite{gururangan2025double} highlighted the importance of including full 4-hole–2-particle (4h2p) correlation effects, along with triply excited configurations for accurately describing doubly ionized states within the EOM-CC framework (DIP-EOM-CCSDT($4h2p$)) using both non-relativistic and spin-free scalar-relativistic Hamiltonians. 

However, for a reliable description of spectral properties and binding energies in transition metal complexes and systems containing heavy elements, it is also necessary to account for both spin-free and spin-dependent relativistic effects. The four-component (4c) Dirac–Coulomb (DC) Hamiltonian offers a rigorous treatment of these effects. Previous studies have used the 4c-DC Hamiltonian to compute DIPs within the two-particle propagator theory\cite{pernpointner2010fourcomponent} and the EOM-CCSD($3h1p$)\cite{pathak2020relativistic} formalism. While the 4c-DC Hamiltonian provides a comprehensive treatment of the relativistic effect, it also comes with a large computational cost. To address this, two-component approaches\cite{hess1986relativistic,lenthe1993relativistic,dyall1997interfacing,nakajima1999new,barysz2001twocomponent,liu2009exact,saue2011relativistic} have gained popularity as efficient alternatives. Among the various two-component relativistic quantum chemistry methods, the exact two-component (X2C)\cite{dyall1997interfacing,liu2009exact,kutzelnigg2005quasirelativistic,ilias2007infiniteorder,dyall2007introduction} method has emerged over the past two decades as a reliable technique to reduce the computational cost of relativistic quantum chemistry calculations, with minimal loss of accuracy. Among the many available variants of X2C Hamiltonian, the X2CAMF method,\cite{liu2018atomic,zhang2022atomic,knecht2022exact} which uses atomic mean-field (AMF) spin–orbit integrals,\cite{hess1996meanfield} stands out for its accuracy and efficiency. It provides a balanced treatment of scalar and spin–orbit relativistic effects with notably reduced computational cost. It has recently been demonstrated that the use of the X2CAMF  Hamiltonian, when combined with a Cholesky decomposition (CD) based treatment of two-electron integrals and a frozen natural spinor (FNS) framework, can lead to highly efficient implementations of relativistic coupled cluster\cite{chamoli2025frozen} and EOM-CC methods.\cite{chamoli2025reduced,mukhopadhyay2025reducedcost} The DIP-EOM-CCSDT($4h2p$) method has recently been extended to both two-component\cite{li2025relativistic} and four-component\cite{mukhopadhyay2025reducedcostfourcomponentrelativisticdouble} relativistic frameworks as well. An alternative approach to calculate the DIP is to use the (0,2) sector of the Fock space multireference coupled cluster method, where the final diagonalization space can be compressed to the determinant space spanned by two hole configurations with respect to the reference states\cite{Musial}. Eliav et al.\cite{Eliav1,Eliav2} have extended the (0,2) sector of the Fock space multi-reference coupled cluster method to the relativistic domain. 

The aim of the present study is to extend the relativistic DIP-ADC method to third order and arrive at an efficient implementation using the CD-X2CAMF approach together with the FNS framework. The structure of this paper is as follows. Section \ref{sec:theory} outlines the theoretical framework, focusing on the relativistic DIP-ADC(3) approach and providing brief overviews of the X2CAMF, CD, and FNS techniques. Section \ref{sec:computn details} outlines the computational details and implementation aspects. In Section \ref{results}, we benchmark the performance of the relativistic DIP-ADC(3) method with available theoretical and experimental results for atoms and molecules. Finally, Section \ref{sec:conclusion} provides our concluding remarks and future perspectives.

\section{\label{sec:theory}Theory}
\subsection{Relativistic DIP-ADC method}
One of the most accurate ways to include relativistic effects in quantum chemistry calculations is to use the four-component Dirac-Coulomb Hamiltonian.\cite{dyall2007introduction} It can be expressed in the occupation-number representation as
\begin{equation}
\hat{H}^{4c} = \sum_{pq} h^{4c}_{pq} \, a^\dagger_p a_q + \frac{1}{4} \sum_{pqrs} g^{4c}_{pqrs} \, a^\dagger_p a^\dagger_q a_s a_r
\label{eq:full4cH}
\end{equation}
Within the no-pair approximation,\cite{sucher1980foundations} the indices $p,q,r,s$ are restricted to positive-energy spinors only. $h^{4c}_{pq}$ and $g^{4c}_{pqrs}$ denote the one-electron integrals and the antisymmetrized two-electron integrals, respectively.

The Algebraic Diagrammatic Construction (ADC) theory was originally formulated using diagrammatic perturbation theory-based expansion for the polarization propagator.\cite{schirmer1982randomphase,vonniessen1984computational,schirmer1983new} An alternative, yet equivalent, approach for deriving the ADC matrix equations is the Intermediate State Representation (ISR).\cite{schirmer2004intermediate,mertins1996algebraic,schirmer1991closedform} To derive the second and third-order relativistic DIP-ADC methods within the ISR framework, we have followed the recipe by Thielen et al.\cite{thielen2023development} The intermediate states (IS) are constructed using the Excited Class Orthogonalization (ECO)\cite{schirmer2018manybody} procedure. In the first step, a correlated excited-state basis, $|\Psi_I^0\rangle$ is generated by applying an excitation operator $\hat{C}_I$ to the correlated ground state, $|\Psi_0\rangle$, where $I$ denotes a compound index that identifies various excitation classes.
\begin{equation}
|\Psi_I^0\rangle = \hat{C}_I |\Psi_0\rangle
\end{equation}The specific form of the $\hat{C}_I$ operator depends on the nature of the target state being investigated. For the DIP case:\\
\begin{equation}
\begin{aligned}
\{\hat{C}_I^{\text{DIP}} \} = \{ c_ic_j, c^\dagger_a c_i c_jc_k, c^\dagger_b c^\dagger_a c_i c_j c_kc_l, \ldots;\\i<j<k \ldots, a< b < \cdots \}.
\end{aligned}
\end{equation}

In general, correlated excited states, $|\Psi^0_I\rangle$ are not orthogonal. To address this, the Gram–Schmidt orthogonalization\cite{mertins1996algebraic} procedure is first applied to generate a set of precursor states, $|\Psi_I^\#\rangle$, which are subsequently subjected to symmetric orthonormalization to construct the final Intermediate State (IS) basis, $|\tilde{\Psi}_I\rangle$.\\
Now, in the ISR formalism, the $k$-th doubly-ionized state is described as follows,
\begin{equation}
|\Psi^{\text{DIP}}_k\rangle = \sum_I Y_{Ik} |\tilde{\Psi}_I^{\text{DIP}}\rangle.
\label{eq:dip_expansion}
\end{equation}
The coefficient matrix $\mathbf{Y}$ and eigenvalues or DIP values, $\boldsymbol{\Omega}$, are obtained from diagonalization of the ADC secular matrix, shifted by the ground state energy ($E_0$).
\begin{equation}
\mathbf{M} \mathbf{Y} = \mathbf{Y} \boldsymbol{\Omega}
\end{equation}
where the matrix elements of $\mathbf{M}$, in the IS basis are given by,
\begin{equation}
\textbf{M}_{IJ} = \langle \tilde{\Psi}_I | \hat{H}^{4c} - E_0 | \tilde{\Psi}_J \rangle
\end{equation}
and 
\begin{equation}
Y_{Ik} = \langle \tilde{\Psi}_I | \Psi_k \rangle; \quad \mathbf{Y}^\dagger \mathbf{Y} = \mathbf{1}
\end{equation}
The resulting secular matrix, $\textbf{M}_{IJ}$, is expanded as a perturbation series, typically expressed as:
\begin{equation}
    \mathbf{M} = \mathbf{M}^{(0)}+\mathbf{M}^{(1)}+\mathbf{M}^{(2)}+\mathbf{M}^{(3)}+\cdots
\end{equation}
Truncating this expansion at a given order $n$, defines the ADC($n$) level of theory. For instance, retaining terms up to second order yields the ADC(2) approximation, while truncation at third order leads to the ADC(3) method. The extended version of strict ADC(2) for DIP, referred to as DIP-ADC(2)-x,\cite{dreuw2015algebraic} introduces an ad hoc first-order expansion of the $3h1p$–$3h1p$ block. All other matrix blocks, however, remain treated according to the strict ADC(2) (ADC(2)-s) scheme. The explicit programmable expressions for the sigma vectors corresponding to the DIP-ADC(2) and DIP-ADC(3) methods are provided in the Supporting Information. To keep the implementation independent of the framework for generating the underlying DHF spinors, the current implementation does not impose the Kramers restriction.\cite{dyall2007introduction}. 
It should be noted that the relativistic third-order ADC(3) implementation for ionization potential, electron affinity, and excitation energy has already been described in the literature.\cite{Sudiptaadc3,sokolov,pernpointner2004one2}

\subsection{The X2CAMF scheme}When using a kinetically balanced basis,\cite{stanton1984kinetic} the two-electron interaction terms in the DC Hamiltonian (Eq.\,\ref{eq:full4cH}) may be separated into spin-free (SF) and spin-dependent (SD) parts via a spin-separation formalism.\cite{dyall1994exact}
\begin{equation}
g^{4c}_{pqrs} = g^{4c,\mathrm{SF}}_{pqrs} + g^{4c,\mathrm{SD}}_{pqrs}
\label{eq:simple}
\end{equation}
The first term in Eq.\,\ref{eq:simple}, representing the SF part of the two-electron interaction, can be approximated by the non-relativistic counterpart, when the effects of the two-electron scalar picture change are neglected. Thus Eq.\,\ref{eq:full4cH} becomes
\begin{equation}
\begin{aligned}
\hat{H}^{4c} \approx  \sum_{pq} h^{4c}_{pq} a_p^\dagger a_q
&+ \frac{1}{4} \sum_{pqrs} g^{\mathrm{NR}}_{pqrs} \, a_p^\dagger a_q^\dagger a_s a_r \\
&+ \frac{1}{4} \sum_{pqrs} g^{4c,\mathrm{SD}}_{pqrs} a_p^\dagger a_q^\dagger a_s a_r \,.
\label{eq:4cH_after_split}
\end{aligned}
\end{equation}
By leveraging the local behavior of spin–orbit coupling, the third term in Eq.\,\ref{eq:4cH_after_split} can be effectively treated using the atomic mean-field (AMF) approximation.\cite{hess1996meanfield}
\begin{equation}
\begin{aligned}
\frac{1}{4} \sum_{pqrs} g^{\mathrm{4c,SD}}_{pqrs} \, a_p^\dagger a_q^\dagger a_s a_r 
&\approx \sum_{pq} g^{\mathrm{4c,AMF}}_{pq} \, a_p^\dagger a_q \\
&= \sum_{pq} \sum_A \sum_i n_{i,A} \, g^{{4c,SD}}_{pi_A\, qi_A} \, a_p^\dagger a_q
\end{aligned}
\end{equation}

Here, $A$ labels individual atoms in the molecule; $i$ denotes a predefined set of occupied spinors, and $n_{i,A}$ denotes the corresponding occupation numbers. Now, the Hamiltonian in Eq.\,\ref{eq:4cH_after_split}, transforms into,
\begin{equation}
\begin{aligned}
\hat{H}^{\mathrm{4c}} \approx 
\sum_{pq} h^{\mathrm{4c}}_{pq} \, a_p^\dagger a_q
&+\frac{1}{4} \sum_{pqrs} g^{\mathrm{NR}}_{pqrs} \, a_p^\dagger a_q^\dagger a_s a_r \\
&+\sum_{pq} g^{\mathrm{4c,AMF}}_{pq} \, a_p^\dagger a_q 
\label{eq:hamil_seperated}
\end{aligned}
\end{equation}
Transforming into the two-component framework using the X2C method\cite{liu2009exact} yields the X2CAMF Hamiltonian, given by
\begin{equation}
\begin{aligned}
\hat{H}^{\text{X2CAMF}} = \sum_{pq} h^{\text{X2C-1e}}_{pq} \, a^\dagger_p a_q
&+ \frac{1}{4} \sum_{pqrs} g^{\text{NR}}_{pqrs} \, a^\dagger_p a^\dagger_q a_s a_r \\
&+ \sum_{pq} g^{\text{2c,AMF}}_{pq} \, a^\dagger_p a_q \,.
\end{aligned}
\end{equation}
It takes the form of a compact one-electron effective operator coupled with the standard non-relativistic two-electron operator 
\begin{equation}
\hat{H}^{\text{X2CAMF}} = \sum_{pq} h^{\text{X2CAMF}}_{pq} \, a^\dagger_p a_q
+ \frac{1}{4} \sum_{pqrs} g^{\text{NR}}_{pqrs} \, a^\dagger_p a^\dagger_q a_s a_r
\label{eq:hx2camf_final}
\end{equation}
with
\begin{equation}
h^{\text{X2CAMF}} = h^{\text{X2C-1e}} + g^{\text{2c,AMF}}
\label{eq:hx2camf_definition}
\end{equation}
The X2CAMF scheme\cite{liu2018atomic,zhang2022atomic,knecht2022exact} circumvents the need to evaluate relativistic two-electron integrals, which makes it particularly advantageous for treating large molecules containing heavy elements with reduced computational expense. A more detailed explanation of the X2CAMF implementation used in the present study can be found in the Ref. \onlinecite{zhang2022atomic}.

\subsection{Cholesky Decomposition}
As originally proposed by Beebe and Linderberg,\cite{beebe1977simplifications} CD provides an efficient way to approximate a positive semi-definite electron repulsion integral (ERI) tensor as a product involving a lower triangular matrix, known as Cholesky vectors (CVs). Within this framework, the symmetric ERI tensor in the atomic orbital (AO) basis can be expressed as:
\begin{equation}
(\mu\nu|k\lambda) \approx \sum_{P}^{n_{\text{CD}}} L^P_{\mu\nu} L^P_{k\lambda}
\end{equation}
Here, $\mu,\nu, k, \lambda$ denote atomic orbital (AO) indices, $L^P_{\mu\nu}$ represents the $P$-th CV, and $n_{CD}$ is the total number of CVs used in the CD. Both single-step and two-step algorithms\cite{aquilante2011cholesky,folkestad2019efficient,zhang2021minimal} have been proposed for performing the CD of ERIs efficiently. In this work, we adopt the conventional single-step algorithm, wherein CVs are generated iteratively by identifying the largest diagonal elements of the ERI matrix, ($\mu\nu|\mu\nu$). The decomposition continues until the largest diagonal element falls below a predefined Cholesky threshold 
$\tau$, which controls the accuracy of the decomposition.
The Cholesky vectors computed in the AO basis are transformed to the MO basis through contraction with the AO–MO transformation matrix obtained from the SCF procedure, as given by:
\begin{equation}
L^{\text{P}}_{pq} = \sum_{\mu\nu} C^*_{\mu p}  L^{\text{P}}_{\mu\nu}  C_{\nu q}
\label{eq:lp_transform}
\end{equation}
Using the transformed CVs, the anti-symmetrized two-electron integrals in the MO basis are as follows:
\begin{equation}
\langle pq || rs \rangle = \sum_P \left( L^P_{pr} \ L^P_{qs} - L^P_{ps} \ L^P_{qr} \right)
\label{eq:antisym_integral_df}
\end{equation}

In our implementation, we explicitly compute and store only those two-electron integrals that involve at most two virtual orbital indices. Integrals of the form $\langle ab||cd \rangle$ and $\langle ab||ci \rangle$ are not stored; instead, they are generated on the fly as required during the computation. This selective treatment significantly reduces memory demands while preserving computational efficiency.

\subsection{Frozen Natural Spinor technique}
DIP-ADC(3) calculations are computationally demanding, primarily due to the large virtual spinor space required to capture electron correlation accurately. Simply removing high-energy canonical virtual spinors often leads to slow convergence and inaccurate results compared to calculations performed with the untruncated basis set. A more effective strategy involves using natural spinors,\cite{chamoli2022reduced} the relativistic analogs of natural orbitals,\cite{lowdin1955quantum} which are obtained by diagonalizing the correlated one-particle reduced density matrix (1-RDM). In this study, we adopt the FNS scheme, in which the occupied orbitals remain unchanged from the SCF calculation, while the virtual space is spanned by the natural spinors generated by diagonalization of the virtual–virtual block of 1-RDM. Among the various flavors of natural spinors available,\cite{chamoli2022reduced,chakraborty2025lowcost,mukhopadhyay2025reducedcost} we have used the MP2 natural spinors.\cite{chamoli2022reduced,surjuse2022lowcost,chamoli2024relativistic,yuan2022assessing} The virtual–virtual block of the 1-RDM in MP2 natural spinors is constructed as follows:
\begin{equation}
D_{ab} = \frac{1}{2} \sum_{ijc} \frac{\langle ac || ij \rangle \langle ij || bc \rangle}{\varepsilon^{ij}_{ac}\ \varepsilon^{ij}_{bc}}
\label{eq:Dab_definition}
\end{equation}
where
\begin{equation}
\varepsilon^{ij}_{ac} = \varepsilon_i + \varepsilon_j - \varepsilon_a - \varepsilon_c
\label{eq:denominator_def}
\end{equation}
where $\varepsilon$  are the molecular spinor energies.
Virtual natural spinors ($V$) are subsequently obtained by diagonalizing the $D_{ab}$, as follows:
\begin{equation}
    D_{ab}V=Vn
\end{equation}
The virtual natural spinor basis $V$ is then truncated using a predefined occupation number threshold, $n_{\text{thresh}}$. All virtual spinors with occupation numbers $n < n_{\text{thresh}}$ are excluded, resulting in a reduced virtual space denoted by $\tilde{V}$.
Subsequently, the virtual–virtual block of the Fock matrix, $F_{VV}$, is projected onto the truncated virtual natural spinor basis,
\begin{equation}
    {\tilde{F}}_{VV}={\tilde{V}}^\dagger F_{VV} {\tilde{V}}
\end{equation}
The resulting ${\tilde{F}}_{VV}$ matrix is then diagonalized to obtain the semi-canonical virtual natural spinors ($\tilde{Z}$) along with their corresponding orbital energies ($\tilde{\varepsilon}$).
\begin{equation}
    {\tilde{F}}_{VV} \tilde{Z}= \tilde{Z}\tilde{\varepsilon}.
\end{equation}
The matrix $B$ serves as the transformation operator that converts the canonical virtual spinor basis into the semi-canonical natural virtual spinor basis.
\begin{equation}
    B= \tilde{V} \tilde{Z}.
\end{equation}
Hence, the FNS basis is constructed from canonical occupied spinors and semi-canonicalized, truncated virtual natural spinors. 
\begin{table}
\caption{\label{tab:compare-4c-cd}Comparison of DIP values (in eV) for selected atoms and molecules, calculated using the 4c-FNS-DIP-ADC(3) and FNS-CD-X2CAMF-DIP-ADC(3) methods with the dyall.av4z basis set.}
\begin{ruledtabular}
\begin{tabular}{cccc}
Atom/Molecule & State & 4c-FNS & FNS-CD-X2CAMF \\
\hline
\multirow{5}{*}{Kr}
& $^3P_2$ & 38.341 & 38.341 \\
& $^3P_1$ & 38.917 & 38.917 \\
& $^3P_0$ & 39.018 & 39.017 \\
& $^1D_2$ & 40.218 & 40.218 \\
& $^1S_0$ & 42.577 & 42.577 \\
\hline
\multirow{5}{*}{Xe}
& $^3P_2$ & 32.981 & 32.981 \\
& $^3P_1$ & 34.200 & 34.199 \\
& $^3P_0$ & 34.020 & 34.020 \\
& $^1D_2$ & 35.165 & 35.164 \\
& $^1S_0$ & 37.581 & 37.580 \\
\hline
\multirow{4}{*}{Cl$_2$}
& $X^3\Sigma^-$ & 31.272 & 31.272 \\
& $a^1\Delta$   & 31.773 & 31.774 \\
& $b^1\Sigma^+$ & 32.157 & 32.158 \\
& $c^1\Sigma^-$ & 33.170 & 33.171 \\
\hline
\multirow{3}{*}{HBr}
& $X^3\Sigma^-$ & 32.810 & 32.810 \\
& $a^1\Delta$   & 34.212 & 34.213 \\
& $b^1\Sigma^+$ & 35.524 & 35.525 \\
\end{tabular}
\end{ruledtabular}
\end{table}
\section{\label{sec:computn details}Computational details}
The relativistic DIP-ADC methods up to third order have been implemented in our in-house quantum chemistry package \texttt{BAGH}.\cite{dutta2025bagh} While the core architecture of \texttt{BAGH} is written in Python, computationally intensive routines are handled via Fortran and Cython for efficiency. It supports a range of non-relativistic and relativistic (4c and 2c) electronic structure 
\begin{figure}
    \centering
    \includegraphics[width=\columnwidth]{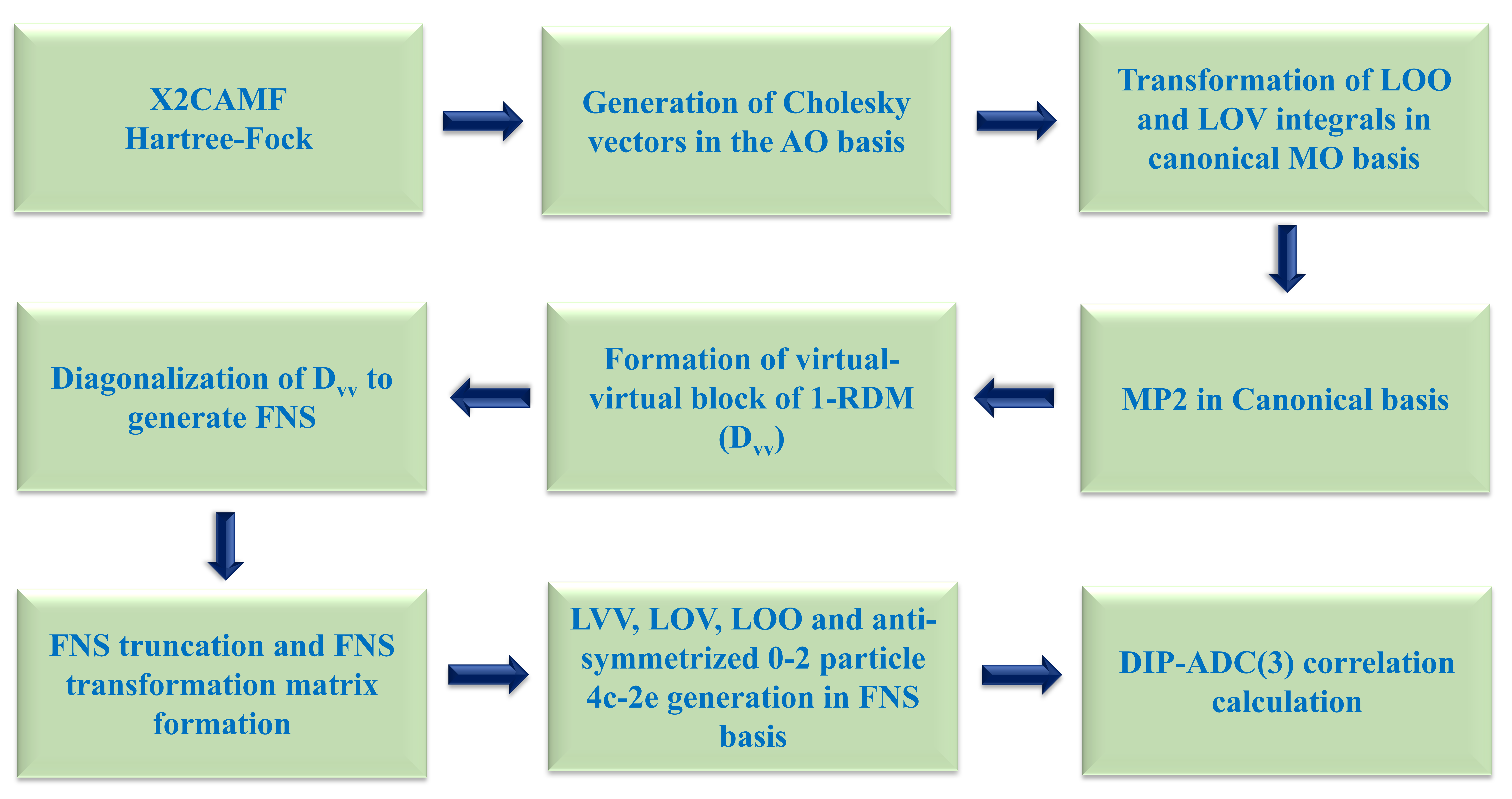}
    \caption{\label{fig:algo} Schematic representation of the FNS-CD-X2CAMF-DIP-ADC(3) algorithm.}
\end{figure}
methods, including many-body perturbation theory, coupled-cluster, and ADC approaches. \texttt{BAGH} utilizes external quantum chemistry programs for the generation of one and two-electron integrals, and presently supports interfaces with \texttt{PySCF},\cite{sun2015libcint,sun2018pyscf,sun2020recent} \texttt{GAMESS-US},\cite{barca2020recent} \texttt{DIRAC},\cite{jensen2022dirac} and \texttt{socutils}.\cite{wang2025socutils} In this work, all calculations were carried out using the \texttt{PySCF} interface. The X2CAMF-HF calculations were carried out using the \texttt{socutils} package. Cholesky vectors of LOO and LOV type are generated prior to the construction of the FNS basis and stored for subsequent use, where O and V
denote occupied and virtual spinors, respectively. The LVV-type Cholesky vectors are only generated in the FNS basis.  In all calculations, the Cholesky vectors were generated using a fixed decomposition threshold of $10^{-5}$, which was found to give similar accuracy as that of the standard integrals.\cite{chamoli2025frozen,chamoli2025reduced} The FNS basis is then constructed from the ground-state MP2 amplitudes. Following this, LVV-type Cholesky vectors are generated directly in the FNS basis. This approach not only speeds up the integral transformation step but also greatly reduces memory usage. Finally, the DIP-ADC(3) equations in the FNS basis are solved using a Davidson iterative diagonalization scheme\cite{DAVIDSON197587}, with initial guess vectors obtained from DIP-ADC(1) calculations. Figure \ref{fig:algo} provides a schematic overview of the FNS-CD-X2CAMF-DIP-ADC(3) algorithm. In the four-component and non-relativistic (NR) calculations, the CD approximation to the two-electron integrals was not employed. The dyall.avxz (x = 2, 3, 4) basis sets\cite{dyall1998relativistic,dyall2002relativistic,dyall2006relativistic,dyall2016relativistic} were employed for the valence DIP calculations. Diffused versions of the dyall.av4z basis set were generated by augmenting with single and double sets of diffuse functions, generated using the \texttt{DIRAC}\cite{jensen2022dirac} software package. Experimental geometries of the diatomic molecules were obtained from the NIST database.\cite{johnson2022nist} Throughout this work, the frozen core approximation was employed for all correlation calculations.

\section{\label{results}Results and discussions}
\subsection{\label{subsec:choice fns_thresh}Choice of FNS threshold} The formation of doubly ionized states involves a large orbital 
\begin{figure}
    \centering
    \includegraphics[width=\columnwidth]{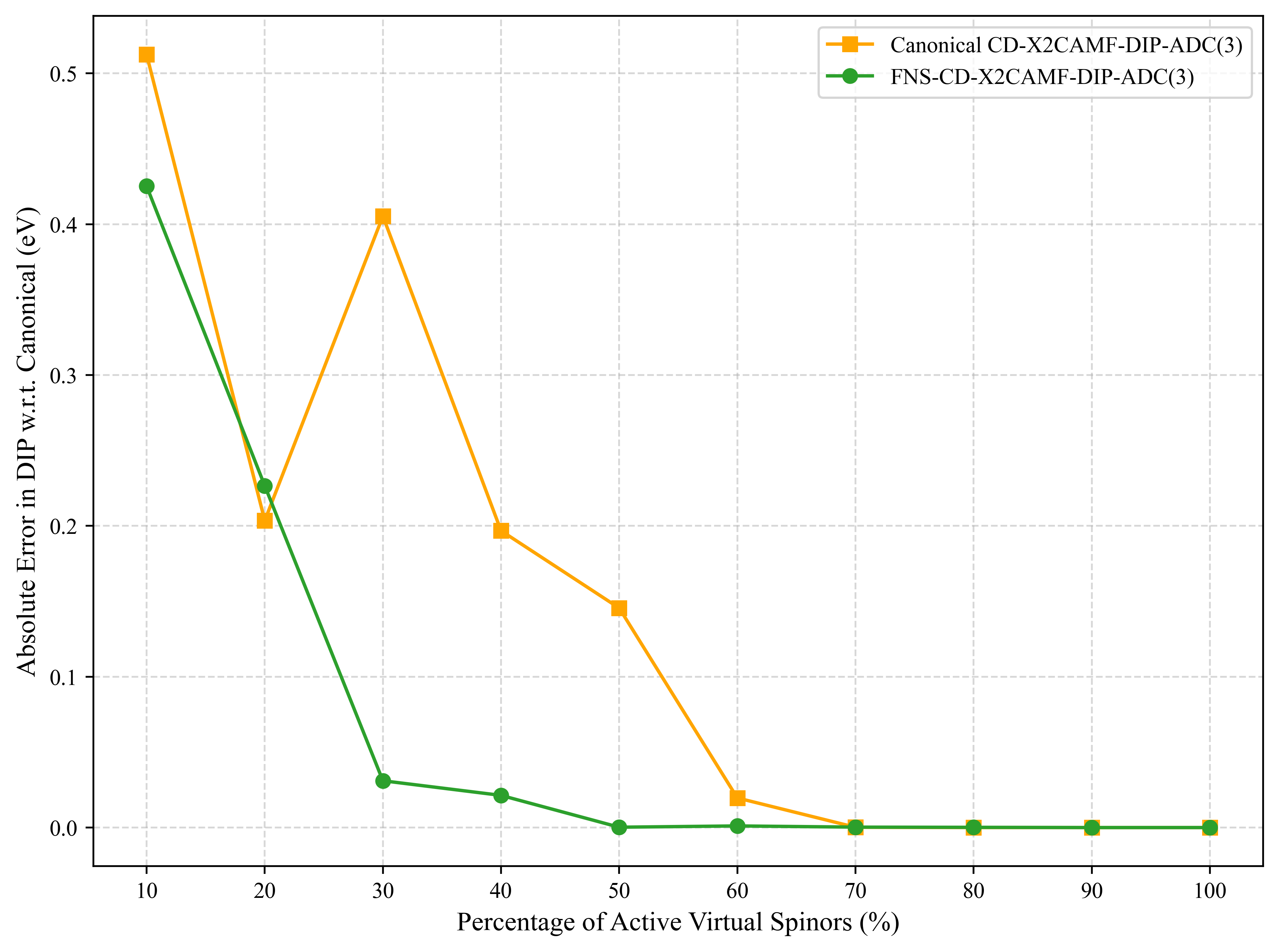}
    \caption{\label{fig:povo_plot} Convergence of the absolute error (in eV) in the lowest DIP value with respect to the size of the virtual space in the canonical and FNS versions of CD-X2CAMF-DIP-ADC(3) method for the Cl$_2$ molecule. The full basis canonical CD-X2CAMF-DIP-ADC(3) result is used as reference, and the dyall.av4z basis set has been used for the calculations.}
\end{figure}
relaxation effect. Therefore, the FNS-induced truncation of $3h1p$ configurations in DIP-ADC(3) can significantly affect the accuracy of the calculation. It is essential to
\begin{figure}
    \centering
    \includegraphics[width=\columnwidth]{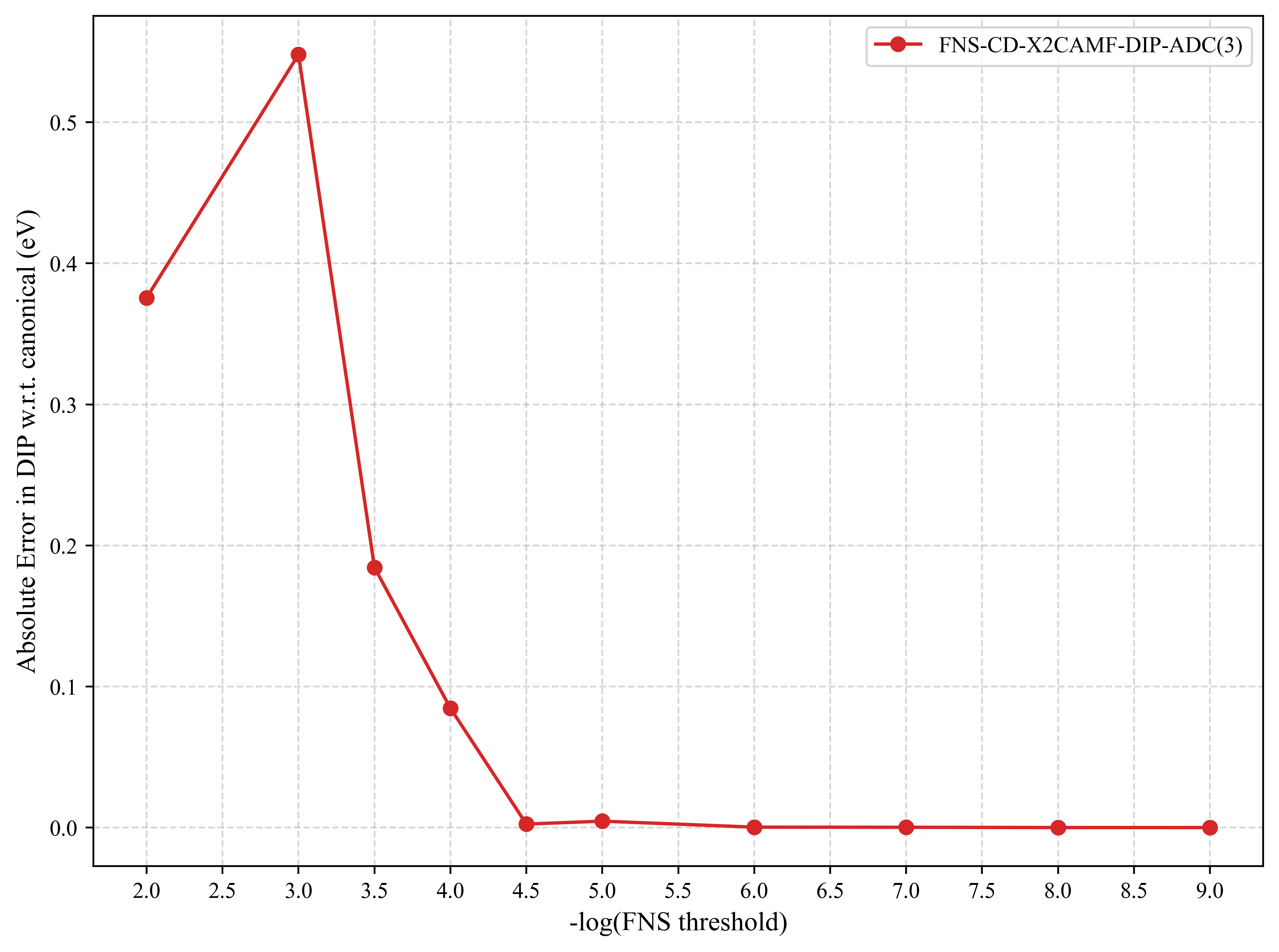}
    \caption{\label{fig:fnsthresh} Convergence of the absolute error in the DIP value (in eV) computed using FNS-CD-X2CAMF-DIP-ADC(3) method, with respect to the FNS threshold
for Cl$_2$ molecule. The dyall.av4z basis set has been used for the calculations, and the untruncated canonical value is taken as the reference.}
\end{figure}
\begin{table*}[t]
    \caption{\label{tab:basis-set-conv}Basis set convergence of FNS-CD-X2CAMF-DIP-ADC(3) values (in eV) of Cl$_2$ molecule in different Dyall basis sets.}
    \begin{ruledtabular}
    \begin{tabular}{ccccccc}
    State & dyall.av2z & dyall.av3z & dyall.av4z & s-aug-dyall.av4z & d-aug-dyall.av4z & Expt.\cite{mcconkey1994threshold} \\
    \hline
    $X^3\Sigma^-$  & 30.90 & 31.11 & 31.27 & 31.27 & 31.28 & 31.13 \\
    $a^1\Delta$    & 31.44 & 31.62 & 31.77 & 31.77 & 31.78 & 31.74 \\
    $b^1\Sigma^+$  & 31.80 & 32.00 & 32.16 & 32.16 & 32.17 & 32.12 \\
    $c^1\Sigma^-$  & 32.82 & 33.01 & 33.17 & 33.17 & 33.18 & 32.97 \\
    \end{tabular}
    \end{ruledtabular}
\end{table*}
benchmark the effect of the truncation on the calculated DIP values. Figure \ref{fig:povo_plot} shows the absolute error in DIP values, as a function of the percentage of virtual orbitals retained in the correlation treatment. The untruncated X2CAMF Hamiltonian-based DIP-ADC(3) canonical value is used as the reference. The data correspond to the lowest vertical DIP of the Cl$_2$ molecule computed with the dyall.av4z basis set. The results clearly show that the FNS approach converges much more rapidly than that obtained by truncating canonical virtual spinors. Even with only 40\% of the frozen natural virtual spinors retained, the FNS-DIP-ADC(3) method yields an absolute error of less than 0.05 eV, whereas the same level of truncation in the canonical basis results in a much larger error of approximately 0.2 eV. Convergence is achieved with just 50\% of the virtual spinors in the FNS basis, compared to the 70\% required in the canonical case. A more convenient and systematic way to truncate the FNS basis is to apply an occupation number threshold, referred to here as the FNS threshold, which excludes all spinors with occupation numbers below a specified value. In Figure \ref{fig:fnsthresh}, we present the absolute error in the DIP energy, computed using the FNS-DIP-ADC(3) method, relative to its canonical counterpart. The error is plotted as a function of the negative logarithm of the FNS threshold for the ground state of the doubly ionized Cl$_2$ molecule, using the dyall.av4z basis set. It can be observed that at a threshold of $10^{-4.5}$, the FNS-DIP-ADC(3) result almost converges to the canonical
value.  Therefore, all subsequent calculations are performed using this threshold.
\subsection{\label{subsec:bench CD against 4c}Benchmarking FNS-CD-X2CAMF results against 4c-DIP-ADC(3) method}
To evaluate the performance of the FNS-CD-X2CAMF-based method against the four-component FNS approach, we computed vertical DIPs for Kr, Xe, Cl$_2$, and HBr using the DIP-ADC(3) method in both frameworks with the dyall.av4z basis set. The results are presented in Table \ref{tab:compare-4c-cd}. The results clearly show that the FNS-CD-X2CAMF-DIP-ADC(3) method yields values nearly identical to the four-component Dirac-Coulomb results across all systems and states considered, with error not exceeding 0.001 eV. This confirms that the former approach can reproduce four-component relativistic results with negligible error, while offering substantial gains in computational speed and memory efficiency. Consequently, all subsequent DIP calculations in this work employ the FNS-CD-X2CAMF-DIP-ADC methods.

\subsection{\label{subsec:basis_conv}Choice of basis set}
\begin{table}
\caption{\label{tab:diperrorinert}
Comparison of error in DIP values (in eV) using 4c-FNS-CCSD($3h1p$), FNS-CD-X2CAMF-ADC(2), and FNS-CD-X2CAMF-ADC(3) methods with experimental values for Ar, Kr, Xe, and Rn in the dyall.av4z basis set.}
\begin{ruledtabular}
\begin{tabular}{cccccc}
Atom & State & CCSD(3h1p)\cite{mukhopadhyay2025reducedcostfourcomponentrelativisticdouble} & ADC(2) & ADC(3) & Expt.\cite{kramida2024nist} \\
\hline
\multirow{5}{*}{Ar} 
& $^3P_2$ & 0.24 & -2.62 & 0.07 & 43.39 \\
& $^3P_1$ & 0.24 & -2.63 & 0.08 & 43.53 \\
& $^3P_0$ & 0.25 & -2.63 & 0.08 & 43.58 \\
& $^1D_2$ & 0.27 & -2.77 & 0.14 & 45.13 \\
& $^1S_0$ & 0.35 & -2.61 & 0.23 & 47.51 \\
\hline
\multirow{5}{*}{Kr} 
& $^3P_2$ & 0.14 & -2.14 & -0.02 & 38.36 \\
& $^3P_1$ & 0.16 & -2.20 & 0.00  & 38.92 \\
& $^3P_0$ & 0.18 & -2.20 & 0.00  & 39.02 \\
& $^1D_2$ & 0.18 & -2.29 & 0.04  & 40.18 \\
& $^1S_0$ & 0.23 & -2.21 & 0.11  & 42.46 \\
\hline
\multirow{5}{*}{Xe} 
& $^3P_2$ & 0.06 & -1.73 & -0.12 & 33.11 \\
& $^3P_1$ & 0.06 & -1.86 & -0.12 & 34.32 \\
& $^3P_0$ & 0.10 & -1.74 & 0.09  & 34.11 \\
& $^1D_2$ & 0.10 & -1.89 & -0.06 & 35.23 \\
& $^1S_0$ & 0.14 & -1.87 & 0.00  & 37.58 \\
\hline
\multirow{1}{*}{Rn} 
& $^3P_2$ & 0.02 & -1.39 & -0.43 & 29.74 \\
\hline
MAE  &   --  & 0.17 & 2.17 & 0.10 & --\\
MAD  &   --  & 0.35 & 2.77 & 0.43 & --\\
STD  &   --  & 0.09 & 0.39 & 0.14 & --\\
RMSD &   --  & 0.19 & 2.21 & 0.14 & --\\
\end{tabular}
\end{ruledtabular}
\end{table}
To understand the effect of basis sets on the calculated DIP values, we utilize the dyall.avxz (x = 2, 3, 4) series,\cite{dyall1998relativistic,dyall2002relativistic,dyall2006relativistic,dyall2016relativistic} which provides a systematically improvable basis set hierarchy. Additionally, to examine the role of diffuse functions, we consider both singly and doubly augmented versions of the dyall.av4z basis set. The results, presented in Table~\ref{tab:basis-set-conv}, illustrate the dependency of DIP values in different basis sets for $X\:^{3}\Sigma^{-}$, $a\:^{1}\Delta$,
$b\:^{1}\Sigma^{+}$, and $c\:^{1}\Sigma^{-}$
states of the \Cltwoplus molecule. It is evident that increasing the basis set quality from double-zeta to triple-zeta leads to an average increase of approximately 0.2 eV in the DIP values for the 
\begin{table}
\caption{\label{tab:diperror_diatomics}
Comparison of error in DIP values (in eV) for Cl$_2$, Br$_2$, HBr, and HI molecules using 4c-FNS-CCSD($3h1p$), FNS-CD-X2CAMF-ADC(2), and FNS-CD-X2CAMF-ADC(3) methods with experimental values in dyall.av4z basis set.}
\begin{ruledtabular}
\begin{tabular}{cccccc}
Molecule & State & CCSD(3h1p)\cite{mukhopadhyay2025reducedcostfourcomponentrelativisticdouble} & ADC(2) & ADC(3) & Expt.\cite{mcconkey1994threshold,fleig2008theoretical,eland2003completeb,yencha2004photodouble}\\
\hline
\multirow{4}{*}{Cl$_2$}
& $X^3\Sigma^-$ & 0.47 & -1.97 & 0.14 & 31.13\\
& $a^1\Delta$    & 0.35 & -1.93 & 0.03 & 31.74 \\
& $b^1\Sigma^+$  & 0.47 & -1.75 & 0.04 & 32.12 \\
& $c^1\Sigma^-$  & 0.54 & -1.14 & 0.20 & 32.97 \\
\hline
\multirow{4}{*}{Br$_2$}
& A $0_g$ & 0.14 & -1.74 & 0.00 & 28.39\\\
& A $1_g$    & 0.13 & -1.78 & 0.02 & 28.53 \\
& A $2_g$  & 0.27 & -1.58 & 0.04 & 28.91 \\
& A $0_g$  & 0.16 & -1.45 & 0.04 & 29.38 \\
\hline
\multirow{3}{*}{HBr}
& $X^3\Sigma^-$   & 0.38 & -1.93 & 0.19 & 32.62\\
& $a^1\Delta$   & 0.43 & -2.02 & 0.26 & 33.95 \\
& $b^1\Sigma^+$   & 0.48 & -1.95 & 0.34 & 35.19 \\
\hline
\multirow{4}{*}{HI}
& $X^3\Sigma_0^-$ & 0.11 & -1.64 & -0.11 & 29.15\\
& $A^3\Sigma_1^-$ & 0.12 & -1.69 & -0.11 & 29.37 \\
& $a^1\Delta$ & 0.15 & -1.71 & -0.03 & 30.39 \\
& $b^1\Sigma^+$ & 0.23 & -1.64 & 0.03  & 31.64 \\
\hline
MAE  & -- & 0.30 & 1.73 & 0.11 & -- \\
MAD  & -- & 0.54 & 2.02 & 0.34 & -- \\
STD  & -- & 0.15 & 0.22 & 0.12 & -- \\
RMSD & -- & 0.33 & 1.74 & 0.14 & -- \\
\end{tabular}
\end{ruledtabular}
\end{table}
four considered states. A further improvement from triple-zeta to quadruple-zeta results in a slightly smaller average increase of 0.16 eV. Unfortunately, quintuple-zeta dyall basis sets are not yet available for Cl.\cite{reitsma2025relativistic} Interestingly, the addition of diffuse functions appears to have minimal impact: the singly augmented dyall.av4z basis set yields no noticeable change in DIP values, while the doubly augmented version results in a modest increase of only
\begin{table}
\caption{\label{tab:br2_dip_methods}
Comparison of DIP values (in eV) of Br$_2$ with previous theoretical and experimental results.}
\begin{ruledtabular}
\begin{tabular}{cccccc}
State & CCSD(3h1p)$^a$& CCSD(3h1p)$^b$ & MRCI\cite{fleig2008theoretical} & ADC(3)$^c$ & Expt.\cite{fleig2008theoretical} \\
\hline
A $0_g$ & --    & 28.53 & 28.39 & 28.39 & 28.39 \\
A $1_g$ & 28.47 & 28.66 & 28.54 & 28.55 & 28.53 \\
A $2_g$ & 29.04 & 29.18 & 29.01 & 28.95 & 28.91 \\
A $0_g$ & 29.52 & 29.54 & 29.45 & 29.42 & 29.38 \\
B $0_u$ & --    & 29.88 & 29.78 & 29.72 & --    \\
B $3_u$ & --    & 29.91 & 29.81 & 29.77 & --    \\
B $2_u$ & 29.79 & 30.26 & 30.16 & 30.12 & 30.30 \\
B $1_u$ & --    & 30.33 & 30.24 & 30.19 & --    \\
B $0_u$ & --    & 30.60 & 30.50 & 30.45 & --    \\
B $1_u$ & --    & 30.62 & 30.52 & 30.48 & --    \\
\end{tabular}
\end{ruledtabular}
\footnotetext[1]{Calculated at dyall.av3z basis. Taken from Pathak et al.\cite{pathak2020relativistic}}
\footnotetext[2]{Calculated at dyall.av4z basis using 4c-FNS approximation. Taken from Mukhopadhyay et al.\cite{mukhopadhyay2025reducedcostfourcomponentrelativisticdouble}}
\footnotetext[3]{FNS-CD-X2CAMF approximation is used.}
\end{table}
0.01 eV per state. Thus, all further computations are performed using the dyall.av4z basis set, without adding any diffuse functions.

\subsection{\label{subsec:compare with exp-theo res}Comparison of DIP-ADC results with available theoretical and experimental data}
To benchmark the performance of  DIP-ADC(2), and DIP-ADC(3) methods, we have computed the DIPs of Ar, Kr, Xe, and Rn, for which corresponding experimental data\cite{kramida2024nist} are available. The results are presented in Table~\ref{tab:diperrorinert}, alongside 4c-FNS-DIP-EOM-CCSD($3h1p$) values and various statistical error metrics, with experimental data taken as the reference. For simplicity, we will hereafter denote DIP-EOM-CCSD($3h1p$) as CCSD($3h1p$) throughout this work. The statistical parameters reported include the mean absolute error (MAE), maximum absolute deviation (MAD), standard deviation (STD), and root mean square deviation (RMSD). It can be clearly seen that DIP-ADC(2) consistently underestimates the double ionization potentials for all the noble gas states studied, with a MAE of 2.17 eV. The largest deviation, 2.77 eV, is observed for the $^1D_2$ state of the Ar atom. This poor performance is not unexpected, as double ionization leads to a significant change in the reference ground state wave function, which results in a significant orbital relaxation effect in the doubly ionized states. A more precise treatment of the $3h1p$ excitations, achieved by incorporating higher-order terms in the ADC secular matrix, is essential for accurately capturing these effects. Since ADC(2) only includes zeroth-order terms in the $3h1p$-$3h1p$ block, it fails to account for the necessary relaxation, resulting in large errors. The first level of improvement is seen with the ADC(2)-x method, which includes first-order corrections in the $3h1p$-$3h1p$ block in an ad hoc manner. This lowers the MAE from 2.17 eV, observed in strict ADC(2), to 0.58 eV in ADC(2)-x (see Table S3 of the Supplementary Material). Among all the methods tested, DIP-ADC(3) shows the best agreement with experiment, with an MAE of only 0.1 eV and a MAD of 0.43 eV. In comparison, the 4c-FNS-CCSD($3h1p$) method yields a MAE of 0.17 eV and a MAD of 0.35 eV.

Table \ref{tab:diperror_diatomics} presents the DIP values for the linear diatomic molecules-Cl$_2$, Br$_2$, HBr, and HI. In the ADC(2) method, a large MAE of 1.73 eV and a MAD of 2.02 eV are observed. In contrast, the significantly lower MAE of 0.30 eV obtained from the FNS-CCSD(3h1p) method highlights the importance of incorporating higher-order treatment of $3h1p$ excitations. This is further supported by the DIP-ADC(3) results, which demonstrate excellent performance with a MAE of just 0.11 eV and an RMSD of 0.14 eV across the 15 states examined in this study. The DIP-ADC(2)-x method yields a significantly lower MAE of 0.55 eV compared to the 1.73 eV observed with the strict ADC(2) method (See Table S4 of the Supplementary Material).

 The Br$_2$ molecule requires special attention. Using the DIP-ADC(3) method, we have computed the ten lowest-energy states of Br$_2^{2+}$ and compared the results (see Table \ref{tab:br2_dip_methods}) with previously reported theoretical and experimental data.\cite{fleig2008theoretical} For comparison, we additionally present results obtained using 4c-CCSD($3h1p$)\cite{pathak2020relativistic} and 4c-FNS-CCSD($3h1p$) method.\cite{mukhopadhyay2025reducedcostfourcomponentrelativisticdouble} Due to the explicit inclusion of spin–orbit coupling in the Hamiltonian, the DIP-ADC(3) method yields excellent agreement with the experimental and MRCI results, even better than the 4c-FNS-CCSD($3h1p$) results. The inclusion of spin-orbit coupling in the calculation is important for reproducing the fine splitting. For example, the splitting between A $0_g$ and A $1_g$ states can not be captured by the scalar-relativistic Hamiltonian.\cite{gururangan2025double} The FNS-CD-X2CAMF-DIP-ADC(3) values for the higher-lying states are also in good
 agreement with available experimental results, as can be seen from Table \ref{tab:br2_dip_methods}.

\subsection{\label{subsec:diff rel hamilton}Different levels of relativistic hamiltonian treatment}
\begin{table}[h]
\caption{\label{tab:dcgb_atoms}Comparison of FNS-CD-X2CAMF-DIP-ADC(3)/dyall.av4z results (in eV) for Ar, Kr, Xe, and Rn atoms across non-relativistic and various relativistic Hamiltonians.}
\begin{ruledtabular}
\begin{tabular}{ccccccc}
Atom & State & NR & DC & DCG & DCB & Expt.\cite{kramida2024nist} \\
\hline
\multirow{5}{*}{Ar}
& $^3P_2$ & 43.56 & 43.46 & 43.45 & 43.45 & 43.39 \\
& $^3P_1$ & 43.56 & 43.60 & 43.59 & 43.59 & 43.53 \\
& $^3P_0$ & 43.56 & 43.66 & 43.65 & 43.65 & 43.58 \\
& $^1D_2$ & 45.28 & 45.26 & 45.25 & 45.25 & 45.13 \\
& $^1S_0$ & 47.75 & 47.74 & 47.73 & 47.73 & 47.51 \\
\hline
\multirow{5}{*}{Kr}
& $^3P_2$ & 38.65 & 38.34 & 38.33 & 38.33 & 38.36 \\
& $^3P_1$ & 38.65 & 38.92 & 38.89 & 38.90 & 38.92 \\
& $^3P_0$ & 38.65 & 39.02 & 38.99 & 39.00 & 39.02 \\
& $^1D_2$ & 40.12 & 40.22 & 40.19 & 40.20 & 40.18 \\
& $^1S_0$ & 42.37 & 42.58 & 42.55 & 42.55 & 42.46 \\
\hline
\multirow{5}{*}{Xe}
& $^3P_2$ & 33.60 & 32.98 & 32.97 & 32.97 & 33.11 \\
& $^3P_1$ & 33.60 & 34.20 & 34.17 & 34.17 & 34.32 \\
& $^3P_0$ & 33.60 & 34.02 & 34.00 & 34.00 & 34.11 \\
& $^1D_2$ & 34.78 & 35.16 & 35.13 & 34.14 & 35.23 \\
& $^1S_0$ & 36.70 & 37.58 & 37.54 & 37.54 & 37.58 \\
\hline
\multirow{1}{*}{Rn}
& $^3P_2$ & 31.14 & 29.31 & 29.29 & 29.30 & 29.74 \\
\hline
MAE  &  --    & 0.38 & 0.10 & 0.11 & 0.10 & -- \\
MAD  &  --    & 1.40 & 0.43 & 0.45 & 0.44 & -- \\
STD  &  --    & 0.52 & 0.14 & 0.15 & 0.14 & -- \\
RMSD &  --    & 0.52 & 0.14 & 0.15 & 0.15 & -- \\
\end{tabular}
\end{ruledtabular}
\end{table}

\begin{table}[h]
\caption{\label{tab:dcgb_molecules}Comparison of FNS-CD-X2CAMF-DIP-ADC(3)/dyall.av4z results (in eV) for Cl$_2$, Br$_2$, HBr, and HI molecules across non-relativistic and various relativistic Hamiltonians.}
\begin{ruledtabular}
\begin{tabular}{ccccccc}
Molecule & State & NR & DC & DCG & DCB & Expt.\cite{mcconkey1994threshold,fleig2008theoretical,eland2003completeb,yencha2004photodouble} \\
\hline
\multirow{4}{*}{Cl$_2$}
& $X^3\Sigma^-$ & 31.32 & 31.27 & 31.27 & 31.27 & 31.13 \\
& $a^1\Delta$    & 31.81 & 31.77 & 31.77 & 31.77 & 31.74 \\
& $b^1\Sigma^+$  & 32.18 & 32.16 & 32.15 & 32.15 & 32.12 \\
& $c^1\Sigma^-$  & 33.24 & 33.17 & 33.17 & 33.17 & 32.97 \\
\hline
\multirow{4}{*}{Br$_2$}
& A $0_g$        & 28.60 & 28.39 & 28.38 & 28.38 & 28.39 \\
& A $1_g$        & 28.60 & 28.55 & 28.53 & 28.54 & 28.53 \\
& A $2_g$        & 29.01 & 28.95 & 28.94 & 28.94 & 28.91 \\
& A $0_g$        & 29.33 & 29.42 & 29.40 & 29.41 & 29.38 \\
\hline
\multirow{3}{*}{HBr}
& $X^3\Sigma^-$  & 32.89 & 32.81 & 32.80 & 32.80 & 32.62 \\
& $a^1\Delta$    & 34.24 & 34.21 & 34.20 & 34.20 & 33.95 \\
& $b^1\Sigma^+$  & 35.49 & 35.53 & 35.51 & 35.51 & 35.19 \\
\hline
\multirow{4}{*}{HI}
& $X^3\Sigma_0^-$ & 29.26 & 29.04 & 29.03 & 29.03 & 29.15 \\
& $A^3\Sigma_1^-$ & 29.26 & 29.27 & 29.25 & 29.25 & 29.37 \\
& $a^1\Delta$     & 30.37 & 30.36 & 30.34 & 30.34 & 30.39 \\
& $b^1\Sigma^+$   & 31.47 & 31.67 & 31.65 & 31.65 & 31.64 \\
\hline
MAE  & -- & 0.15 & 0.11 & 0.10 & 0.10 & -- \\
MAD  & -- & 0.30 & 0.34 & 0.32 & 0.32 & -- \\
STD  & -- & 0.14 & 0.12 & 0.12 & 0.12 & -- \\
RMSD & -- & 0.18 & 0.14 & 0.14 & 0.14 & -- \\
\end{tabular}
\end{ruledtabular}
\end{table}
We now investigate the consequences of neglecting relativistic effects in DIP calculations, as well as the impact of incorporating higher-order terms in the relativistic Hamiltonians. In Table \ref{tab:dcgb_atoms},  we present the DIP-ADC(3) values computed using non-relativistic, Dirac–Coulomb (DC), Dirac–Coulomb–Gaunt (DCG), and Dirac–Coulomb–Breit (DCB) Hamiltonians for selected noble gas atoms. The Gaunt and the Breit interactions have only been considered in the Hartree-Fock step and have not been included in the subsequent integral transformation. One clear observation is that relativistic treatment is crucial for correctly describing the DIP values of heavier elements. Notably, the $^3P_2$ state of Rn is overestimated by about 1.4 eV in the non-relativistic framework, whereas the DC Hamiltonian brings the error down to just 0.43 eV. A similar trend is seen in the doubly ionized states of Xe. Incorporating the Gaunt correction into the DC Hamiltonian leads to a slight reduction in DIP values across all considered states, with the largest decrease of 0.04 eV observed for the $^1S_0$ state of the Xe atom. The change in the DIP value due to the inclusion of Breit correction is negligible over that obtained with the DCG Hamiltonian. We have also investigated the effect of the treatment of relativity for the diatomic molecules considered in this study. The corresponding results are presented in Table \ref{tab:dcgb_molecules}.  The inclusion of relativistic effects using the DC Hamiltonian reduces the MAE from 0.15 eV in a non-relativistic calculation to 0.11 eV.  The importance of incorporating relativistic effects is more pronounced for molecules containing heavy elements. When the Gaunt correction is added to the DC Hamiltonian, the DIP values either remain unchanged or show a slight decrease, with a maximum shift of just 0.02 eV. The inclusion of the Breit correction produces only marginal changes relative to the DCG Hamiltonian. The statistical error parameters for both DCG and DCB results remain nearly identical to those obtained with the DC Hamiltonian, indicating that the effects of Gaunt and Breit corrections have minimal impact in this context.

\section{Computational efficiency}
A central aim of using CD and X2CAMF approximations is to lower the computational cost of DIP-ADC(3) calculations for large basis sets. As a test case, we calculated the DIP of the HI molecule with the dyall.av3z basis using full four-component- (4c-), FNS-4c-, and FNS-CD-X2CAMF- DIP-ADC(3) methods and compared the timings across different stages of the computation. The calculations were performed 
\begin{figure}[h]
    \centering
    \includegraphics[width=\columnwidth]{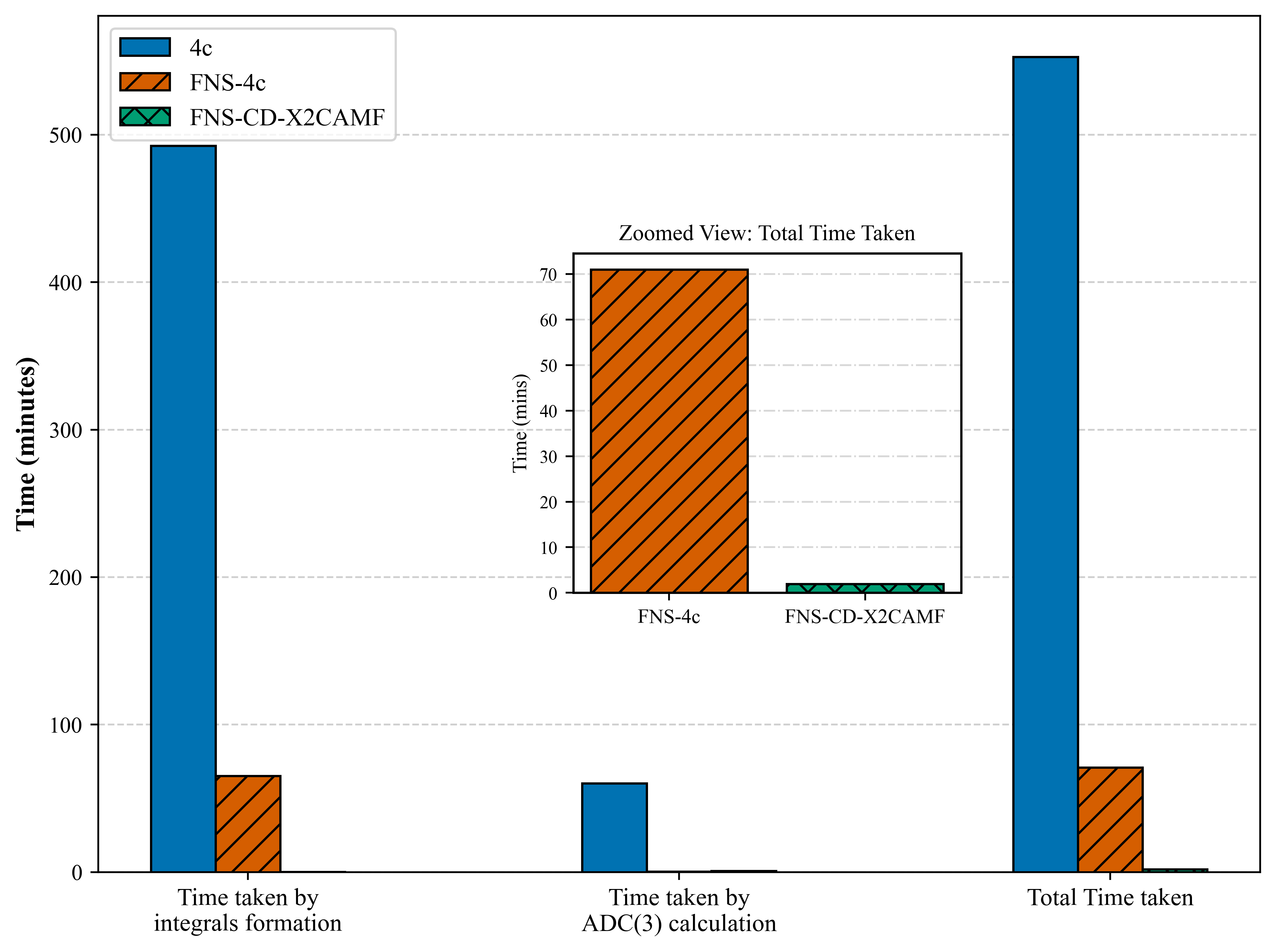}
    \caption{\label{fig:timecal} Comparison of computational timings for different steps in the DIP calculation of HI using canonical 4c, FNS-4c, and FNS-CD-X2CAMF based DIP-ADC(3) methods.}
\end{figure}
using a single thread on a workstation comprising two Intel(R) Xeon(R) Silver 4216 CPUs with a clock speed of 2.10 GHz, and 512 GB of available RAM. All calculations were performed using the frozen-core approximation. For the FNS-based computations, an FNS threshold of $10^{-4.5}$ was used, while a CD threshold of $10^{-5}$ was applied in the FNS-CD-X2CAMF approach. In the full 4c calculation, 382 virtual spinors were correlated, while the FNS procedure retained only 28\% of this space, corresponding to 104 virtual spinors. From Figure \ref{fig:timecal}, it is clear that the integral transformation is the most time-consuming step in both the full 4c and FNS-4c calculations. In the full 4c case, this step required 8 hours, 12 minutes, and 33 seconds, whereas employing the FNS basis reduced the time to 1 hour, 5 minutes, and 16 seconds. The FNS-CD-X2CAMF approach provides further advantages by eliminating the need to generate and store three- and four-virtual integrals, yielding substantial savings in both storage and computational effort. Considering the overall timings, the full 4c method required 9 hours 12 minutes, the FNS-4c approach took about 1 hour 11 minutes, while the FNS-CD-X2CAMF scheme completed the calculation in just 2 minutes. For this particular case, the FNS-CD-X2CAMF method reproduced the full 4c DIP value up to the fourth decimal place in eV, highlighting its outstanding efficiency and its ability to overcome storage limitations.

\section{\label{sec:conclusion}Conclusion}
In summary, we have presented the theory, implementation, and benchmarking of a third-order relativistic ADC method for calculating double ionization potentials. The extension to the third order significantly increases the accuracy over the DIP-ADC(2) method. The computational cost of the relativistic DIP-ADC(3) method can be reduced by using a two-component X2CAMF Hamiltonian combined with the FNS framework and CD-based treatment of the two-electron integrals. The X2CAMF-DIP-ADC(3) method gives almost identical DIP values as those obtained with a four-component DC Hamiltonian. The use of the FNS approximation allows one to truncate the virtual space for the DIP-ADC calculation with systematic controllable accuracy.
Additionally, the use of CD allowed us to employ large basis sets, even for heavy elements. Our analysis shows that lower-order approximations like ADC(2) are not sufficient for reliable predictions of DIP due to their inadequate treatment of the $3h1p$ configuration. The DIPs computed using the DC Hamiltonian at the ADC(3) level of theory showed a MAE of just 0.11 eV when compared with the experimental results, highlighting the robustness and precision of the relativistic ADC(3) method. The results show agreement with experiment that is comparable to, and often better than, the relativistic DIP-EOM-CCSD($3h1p$) method. The inclusion of Gaunt and Breit corrections has a minimal effect on the valence DIP values. To obtain a quantitative agreement for DIP values, one needs to include the $4h2p$ blocks in the calculation, which essentially means going to the ADC(4)\cite{schirmer1983new,leitner2022fourthorder} level of theory. Work is in progress towards developing a lower scaling relativistic DIP-ADC(4) method. 

\section*{SUPPLEMENTARY MATERIAL}
The Supplementary Material provides (a) the working equations for DIP-ADC(3) in algorithmic form, (b) numerical results associated with Figures \ref{fig:povo_plot} and \ref{fig:fnsthresh}, and (c) data from the DIP-ADC(2)-x calculations.

\section*{Acknowledgements}
The authors gratefully acknowledge financial support from IIT Bombay, including the IIT Bombay Seed Grant (Project No. R.D./0517-IRCCSH0-040), CRG (Project No. CRG/2022/005672), and MATRICS (Project No. MTR/2021/000420) projects of DST-SERB; CSIR-India (Project No. 01(3035)/21/EMR-II); DST-Inspire Faculty Fellowship (Project No. DST/INSPIRE/04/2017/001730); and ISRO (Project No. RD/0122-ISROC00-004). The authors also acknowledge the IIT Bombay supercomputing facility and C-DAC resources (Param Smriti, Param Bramha, and Param Rudra) for providing computational time. S.M. gratefully acknowledges support from the Prime Minister’s Research Fellowship (PMRF).

\renewcommand{\refname}{References}
\nocite{apsrev41Control}
\bibliographystyle{aipnum4-1}
%


\end{document}